\documentclass[12pt]{article}

\title{\bf Distinguishing marks of simply-connected  
                 universes}
\author{
M.J. Rebou\c{c}as\thanks{E-mail: reboucas@cbpf.br} \\ 
\\ 
Centro Brasileiro de Pesquisas F\'\i sicas, \\
Rua Dr. Xavier Sigaud 150 \\
22290-180 Rio de Janeiro -- RJ, Brazil
}

\begin{document}
\date{\today}
\maketitle

\begin{abstract} 
A statistical quantity suitable for distinguishing simply-connected 
Robertson-Walker (RW) universes is introduced, and its explicit 
expressions for the three possible classes of simply-connected RW 
universes with an uniform distribution of matter are determined. 
Graphs of the distinguishing mark for each class of RW universes 
are presented and analyzed.
There sprout from our results an improvement on the procedure 
to extract the topological signature of multiply-connected RW 
universes, and a ref\/ined understanding of that topological 
signature of these universes studied in previous works.   
\end{abstract}

\section{Introduction}
\label{intro}
\setcounter{equation}{0}

Current observational data favor the locally homogeneous and isotropic
Friedmann-Lema\^{\i}tre-Robertson-Walker (FLRW) cosmological models as 
approximate descriptions of our universe at least since the recombination 
time. Thus in the framework of the general relativity theory it can be 
described through a Robertson-Walker (RW) metric
\begin{equation}
\label{RWmetric}
ds^2 = dt^2 - R^2(t)\, d \sigma^2 \; ,
\end{equation}
where $t$ is a cosmic time, and 
$d\sigma^2 = d\chi^2 + f^2(\chi)\,[\,d\theta^2+\sin^2\theta\,d\phi^2\,]$ 
with $f(\chi)= \chi\,,\; \sin\chi\,,\; \sinh\chi\,,\;$ depending on the 
sign of the constant spatial curvature ($k = 0, \pm 1$). 
These descriptions, however, are only local and do not f\/ix the global 
shape (topology) of our universe.

Despite the inf\/initely many possibilities for its global topology, 
it is often assumed that spacetime is simply-connected leaving aside 
the hypothesis that the universe may be multiply-connected, and compact
(f\/inite) even in the cases $k=0$ and $k=-1$.
In other words, it is often assumed that the $t=const$ spatial sections 
$M$ of a RW spacetime manifold are one of the following simply-connected 
spaces:  Euclidean $E^3$ ($k=0$), elliptic $S^3$ ($k=1$), or the 
hyperbolic $H^3$ ($k=-1$).  
However, the connectedness (either simply or multiply) for our three-space 
has not been settled by cosmological observations. Thus, the space $M$ 
where we live may also be any one of the possible multiply-connected 
quotient three-spaces $M = \widetilde{M}/\Gamma$, where $\widetilde{M}$ 
stands for $E^3$, $S^3$ or $H^3$, and $\Gamma$ is a discrete group 
of isometries of the covering space $\widetilde{M}$ acting
freely on  $\widetilde{M}$~\cite{Wolf}.

Whether we live in a simply or multiply-connected, f\/inite (compact) 
or inf\/inite (non-compact) space, and what is the size and the shape 
of the universe are open problems modern cosmology seeks to 
solve~\cite{Starkman98}.
The most immediate consequence of multiply-connectedness of the 
universe is that the sky may show multiple images of cosmic 
objects periodically distributed in the space.
This periodic distribution of images arises from the correlations 
in their positions dictated by the discrete isometries of the 
covering group $\Gamma$ of the three-manifold used to model its space
section.

One way to tackle the problems regarding the topology of the 
universe is through a suitable statistical analysis applied to 
catalogs of discrete cosmic sources to f\/ind out whether or not 
there are multiple correlated images of cosmic objects, and eventually
determine the topological features of the universe from the
pattern of images the sky shows.

The correlations among the images of cosmic objects in multiply-%
connected universes can be couched in terms of distance correlations 
between the images.
Indeed, one way of looking for distance correlations between 
cosmic images in multiply-connected universes is by using pair 
separations histograms (PSH), which are functions $\Phi(s_i)$ 
that count the number of pair of images separated by a distance 
that lies in intervals (bins) $J_i$.
The embryonic expectation was that the distance correlations would
manifest as topological spikes in PSH's, and that the spike spectrum 
would be a def\/inite signature of the non-trivial topology~\cite{LeLaLu}.
However, this initial expectation turned out to be false%
~\cite{LeLuUz}~--~\cite{GRT99a}.
Nevertheless, the most striking evidence of multiply-connectedness 
in PSH's is indeed the presence of such topological spikes, which 
arise from translational isometries $g_t \in \Gamma$. 
The non-translational isometries $g_{nt} \in \Gamma$, however, 
manifest as rather tiny deformations of the expected pair separation 
histogram $\Phi^{sc}_{exp}(s_i)$ corresponding to the underlying 
simply-connected universe. 
However, from computer simulations it becomes clear that the expected 
pair separation histogram (EPSH) corresponding to a multiply-connected
universe $\,\Phi_{exp}(s_i)$, which is nothing but an PSH from 
which the statistical noise has been withdrawn, is not a suitable 
quantity for revealing the topology of {\em multiply-connected\/} 
universes~\cite{GRT99b}.

In a recent article, Gomero \emph{et al.\/}~\cite{GRT99b} (see also
~\cite{GRT99b}) have proposed a way of extracting the topological 
signature of any \emph{multiply-connected\/} universe of constant 
curvature by using a \emph{new} quantity $\varphi^{mc}(s_i)\equiv (n-1)\,
[\,\Phi_{exp}(s_i)- \Phi^{sc}_{exp}(s_i)\,]$, where $n$ is the 
number of images. Note, however, that this quantity cannot be 
used as distinguishing marks of \emph{simply-connected} universes 
since it vanishes identically for such universes. 
This amounts to saying that the graphs of $\varphi^{mc}(s_i)$ for all
three classes of RW simply-connected universes which arise from 
(simulated or real) catalogs exhibit nothing but statistical noise, 
and thus $\varphi^{mc}(s_i)$ should not be used as identifying 
markings in the simply-connected cases.
As a matter of fact, the scheme  discussed in~\cite{GRT99b} 
as well as the approaches that make use of the cosmic microwave 
background radiation~\cite{CSS98a}~--~\cite{CW98} were fundamentally 
devised to reveal the possible non-trivial topology of \emph{small} 
universes. However, neither the multiply nor the simply-connectedness 
for our universe has been discarded or conf\/irmed by the current 
astrophysical observations.

In computer-aided simulations the histograms such as the PSH's 
$\Phi(s_i)$ contain statistical f\/luctuations, which can give rise 
to sharp peaks of statistical (non-topological) origin, or can hide 
(or mask) the tiny deformations due to non-translational isometries. 
The most immediate approach to cope with f\/luctuation problems 
in PSH's  is by using the mean pair separation histogram (MPSH)
scheme to obtain the mean PSH $<\!\Phi(s_i)\!>$  rather than 
a single PSH $\,\Phi(s_i)\,$.
In ref.~\cite{GRT99b} they have used the MPSH technique to extract 
the topological signature of RW multiply-connected universes. 
This technique consists in the use of $K$ (say) computer-generated 
comparable catalogs to obtain the mean pair separation histograms 
$<\!\Phi(s_i)\!>\,$ and $<\!\Phi^{sc}(s_i)\!>\,$; and use them as 
approximations for $\Phi_{exp}(s_i)$ and $\Phi^{sc}_{exp}(s_i)\,$, 
to construct the topological signature $\varphi^{mc}(s_i)\simeq (n-1)\,
[<\!\Phi(s_i)\!> - <\!\Phi^{sc}(s_i)\!>]\,$.
Obviously the greater is the number $K$ of catalogs the better are the 
approximations $<\!\Phi(s_i)\!> \,\simeq  \Phi_{exp}(s_i)\,$ 
and $<\!\Phi^{sc}(s_i)\!> \,\simeq \Phi^{sc}_{exp}(s_i)\,$.
  
In this article we point out that the statistical quantity 
$\phi^{sc}(s_i) = \Phi^{sc}_{exp}(s_i)$ is indeed a suitable 
distinguishing mark of the simply-connected RW universes, and 
rederive its explicit expressions for Euclidean, hyperbolic 
and elliptic \emph{simply-connected} universes (spherical balls 
$\mathcal{B}_a$  with radius $a$) fulf\/illed with an uniform 
distribution of cosmic objects. 
In doing so, on the one hand we obtain the exact (free from statistical 
f\/luctuation) expressions for the distinguishing mark 
$\phi^{sc}(s_i)$ of the three possible classes of simply-connected 
RW universes; on the other hand one attains a ref\/ined statistical 
meaning of the signature $\varphi^{mc}(s_i)$ and also obtains an 
improvement on the procedure to extract the topological signature of 
\emph{multiply-connected\/} RW universes devised in ref.~\cite{GRT99b}.

In the next section we set our framework, def\/ine the basic notation, 
and derive the expressions for the distinguishing marks for the three 
possible classes of simply-connected RW universes (Euclidean, hyperbolic, 
elliptic). 
There we also present and analyze graphs of the distinguishing
mark $\phi^{sc}(s_i)$ of simply-connected RW universes, and 
discuss the improvement we have obtained in the procedure to extract 
the topological signature of multiply-connected universes studied in
ref.~\cite{GRT99b}. In the last section we summarize and discuss
our main results and present the concluding remarks.

To close this section a word of clarif\/ication: although throughout 
this article we loosely use the terminology topological signature of a 
universe and/or of a manifold, it should be noted that the topological 
signature actually corresponds to an observed universe, which in this 
paper is a spherical ball $\mathcal{B}_a\subset \widetilde{M}$ of radius 
$a$, which contains the set of the observed images.

%%%
%%%%%%%%
%%%
\section{Distinguishing marks}
\label{sig}
\setcounter{equation}{0}

In this section we will f\/irst set the notation and then recast
in a unif\/ied and compact way the explicit expressions for the 
probability densities obtained in~\cite{BT99} so as to show that 
they can be used as distinguishing mark for the three possible 
classes ($k = 0, \pm 1$) of simply-connected RW universes 
fulf\/illed with an uniform distribution of cosmic objects. 

Let us start by recalling that a \emph{catalog} $\mathcal{C}$ is a 
set of \emph{observed images}, subset of the set $\mathcal{O}$
of \emph{observable} images ($\,\mathcal{C} \subset \mathcal{O}\,$),  
which are clearly contained in the \emph{observable universe}, which 
in turn is the part of the universal covering manifold $\widetilde{M}$ 
causally connected to an image of a given observer. 
The \emph{observed universe} is the part of the observable universe 
which contains all the sources registered in the catalog. 
Our observational limitations are formulated through selection rules 
which dictate how the subset $\mathcal{C}$ arises from $\mathcal{O}$. 
Catalogs whose images obey the same (well-behaved) distribution
law and that follow the same selection rules are said to be 
\emph{comparable catalogs}~\cite{GTRB98}. 
It should be noted that in the process of construction of 
catalogs it is assumed a RW geometry (needed to convert redshift 
into distance) and that a particular type of sources 
(clusters of galaxies, quasars, etc) is chosen from the 
outset. So, for our purpose in the present work, in addition to
the angular positions on the celestial sphere, the relevant 
information registered in a given catalog is the redshift 
corresponding to each image in the catalog. 

Consider a catalog $\mathcal{C}$ with $n$ cosmic images and 
denote by $\eta(s)$ the number of pairs of images whose  
separation is $s$. Consider also that our observed universe is a 
ball of radius $a$ and divide the interval $(0,2a]$ in $m$
equal subintervals $J_i$ of length $\delta s = 2a/m$. Each of
such subintervals has the form
\begin{equation} \label{Ji}
J_i = (s_i - \frac{\delta s}{2} \, , \, s_i + \frac{\delta
s}{2}] \qquad ; \qquad i=1,2, \,\ldots\, ,m \;\, ,
\end{equation}
and is centered at
\begin{displaymath}
s_i = \,(i - \frac{1}{2}) \,\, \delta s \;.
\end{displaymath}
The PSH is a normalized function which counts the number of pair 
of images separated by a distance that lies in the 
subinterval $J_i$. Thus the function PSH is given by
\begin{equation}  \label{defpsh}
\Phi(s_i)=\frac{2}{n(n-1)}\,\,\frac{1}{\delta s}\,
               \sum_{s \in J_i} \eta(s) \;,
\end{equation}
and  is clearly subjected to the normalizing condition
\begin{equation}
\sum_{i=1}^m \Phi(s_i)\,\, \delta s = 1 \; .
\end{equation}

If one considers an ensemble of comparable catalogs%
\footnote{Note that a typical catalog of the ensemble 
ref\/lects (corresponds to) a distribution of images in 
the observed universe $\mathcal{B}_a\,$.}
with the same number $n$ of images, and corresponding to the same 
three-manifold $M$ of constant curvature, one can compute probabilities 
and expected values of quantities which depend on the images in 
the catalogs of the ensemble.
In particular, we can compute the expected number $\eta_{exp}(s_i)$ 
of pairs of cosmic images in a catalog $\mathcal{C}$ of the 
ensemble with separations in $J_i$.
This quantity is quite relevant because from it one has the 
normalized expected pair  separation histogram (EPSH) which 
clearly is given by 
\begin{equation}
\label{defepsh}
\Phi_{exp}(s_i) = \frac{1}{N}\,\,\frac{1}{\delta s}\,\, 
\eta_{exp}(s_i) = \frac{1}{\delta s}\,\,F(s_i) \; ,
\end{equation}
where obviously $N=n(n-1)/2$ is the total number of pairs of cosmic 
images in $\mathcal{C}$, and $F(s_i)=\eta_{exp}(s_i)/N $ is the 
probability that a pair of images be separated by a distance 
that lies in the interval $J_i$.

In what follows we shall consider that we have an ensemble of 
comparable catalogs whose underlying observed universe (spherical 
ball $\mathcal{B}_a$ with radius $a$) are simply-connected and
fulf\/illed with an uniform distribution of pointlike objects. 
We will take $\phi^{sc}(s_i) \equiv \Phi^{sc}_{exp}(s_i)$ as 
distinguishing mark for these three classes of simply-connected 
universes. Clearly for this uniform distribution of objects
all separations $\,0< s_i \leq 2a\,$ are allowed, so the 
identifying markings $\phi^{sc}(s_i)$ are continuous functions 
of $s$ given by 
\begin{equation} \label{epshsc1}
\phi^{sc}(s) = \Phi^{sc}_{exp}\,(s) 
                    = \frac{1}{\delta s}\,F_{sc}(s)\;,
\end{equation} 
where $F_{sc}(s)$ is the probability that a pair of images in a
catalog $\mathcal{C}$, corresponding to a simply-connected 
universe, be separated by a distance $s\,$. For the sake of 
simplicity hereafter we will drop the subscript of $F_{sc}(s)$.

To make explicit that the distinguishing mark depends upon the radius 
of the observed universe we rewrite~(\ref{epshsc1}) in the form
\begin{equation} \label{epshsc2}
\phi^{sc}(a,s) =\Phi^{sc}_{exp}\,(a,s) 
         =\frac{1}{\delta s}\,F(a,s)=\mathcal{F}\,(a,s) \;,
\end{equation} 
where $\mathcal{F}\,(a,s)$ clearly is the probability density, i.e.
it is such that $F(a,s) = \mathcal{F}\,(a,s)\, ds$ gives   
the probability that two arbitrary points in the ball 
$\mathcal{B}_a$ be separated by a distance between $s$ and 
$s+ds$. Equation~(\ref{epshsc2}) makes apparent that the 
$\phi^{sc}(a,s)$ gives essentially the distribution 
of probability for all $s$ in the ball $\mathcal{B}_a$. 
Moreover, since the way one measures the distances varies for 
each constant curvature universe, it is clearly expected 
that the expression for distinguishing mark $\phi^{sc}(a,s)$ 
changes with the three-geometry of these simply-connected universes. 
In what follows we shall recast in a compact way the expressions 
of $\phi^{sc}(a,s)$ for Euclidean, hyperbolic and elliptic 
simply-connected universes~\cite{BT99}. 

Consider in either of the simply-connected three-spaces
a ball $\mathcal{B}_{a}\,$ centered at the origin $O$, and let 
$P$ and $Q$ be two arbitrary points in the ball. Denote by
$r\in [0 ,a]$ the radial position of $P$, and by 
$s\leq 2a$ the distance from $P$ to $Q$ (see f\/igure~1).%
\footnote{In order to encompass the elliptic class in our compact 
approach we shall initially treat only the elliptic cases in which 
the radius $a$ of the universe $\,\mathcal{B}_{a}\subset S^3$ is 
such that $2a<\pi\,R$, where $R$ is the radius of the curvature 
of the geometry, i.e. the scale factor of RW metric~(\ref{RWmetric}) 
for a given time $t=t_0$. Further, for the sake of simplicity and 
without loss of generality we shall also set $R=1$ for both the 
hyperbolic and elliptic cases.}
 
Consider now the quantity $\mathcal{F}(a,r,s)\,dr \,ds\,$, which 
is the probability that $P$ lies in a position between $r$ and $r+dr$, 
times the probability that the separation between $P$ and $Q$ lies 
between $s$ and $s+ds$. 
Clearly for the simply-connected cases we are concerned the 
probability density $\mathcal{F}(a,r,s)$ is proportional to the 
following two areas: 
(i) $\,\mathcal{A}_S(r)$ which is the area of the locus of the 
points $P$ located at a distance $r$ from the origin $0\,$; and 
(ii) the area of the locus of the points $Q$ that are separated 
{}from $P$ by $s$. Note, however, that when $r + s < a$ the latter 
locus is a two-sphere $S^2$ with area $\mathcal{A}_S(s)$, 
whereas when $r+s>a$ it changes into a spherical 
calotte (cap) with area $\mathcal{A}_C(a,r,s)$.

To sum up we have that the expression for the probability density 
$\mathcal{F}(a,r,s)$ for the conf\/iguration in which $P$ is between 
$r$ and $r+dr\,$, and is separated from $Q$ by a distance between $s$ 
and $s+ds$, can be written in the general form
\begin{equation} \label{den}
\mathcal{F}(a,r,s)= \zeta\,\,\mathcal{A}_S(r)\,\left[\,
\mathcal{A}_S(s) \; \Theta (a-s-r) + \mathcal{A}_C(a,r,s)\; 
             \Theta (s+r-a) \,\right]\;, 
\end{equation}
where $\zeta$ is a normalization constant and $\Theta$ is the 
Heaviside function. Obviously, due to the cut-of\/f ef\/fects 
of the function $\Theta$ the f\/irst term of the sum in the
right-hand side of~(\ref{den}) is nonzero only for $r + s < a$, 
whereas the second term is non-null only when $r+s>a\,$.

Now since the area $\mathcal{A}_S(r)\,$ of the two-sphere as well 
as the area of the spherical calotte $\mathcal{A}_C(a,r,s)\,$
depend on what is the simply-connected three-space where the ball 
$\mathcal{B}_{a}\,$ is considered, then to obtain $\mathcal{F}(a,r,s)$ 
for each class of simply-connected universes we are interested one 
ought to: 
(i) calculate the areas $\mathcal{A}_S(r)\,$ and $\mathcal{A}_C(a,r,s)\,$; 
(ii) insert in~(\ref{den}) and integrate $\mathcal{F}(a,r,s)$ from $r=0$ 
to $r=a$; and 
(iii) impose the normalization condition
\begin{equation} \label{normcond}
\int_{0}^{2a} \mathcal{F}(a,s)\,ds = \int_{0}^{2a} 
                        \phi^{sc}(a,s)\,ds=1\;,
\end{equation}
to obtain the value of the normalization constant $\zeta$.
In what follows we shall use this systematic scheme to
determine $\mathcal{F}(a,r,s)$ for the three classes
of simply-connected universes we are concerned.

For Euclidean universe ($\,\mathcal{B}_a \subset E^3$) one obviously
has $\mathcal{A}_S(r)= 4\pi r^2$, and straightforward calculations 
furnish $\mathcal{A}_C(a,r,s)=(\pi s/r)\,[\,a^2-(s-r)^2\,]$. According
to the above-outlined scheme inserting these expressions in~(\ref{den}), 
integrating $\mathcal{F}(a,r,s)$ from $r=0$ to $r=a$, and using the 
condition~(\ref{normcond}) together with~(\ref{epshsc2}), one obtains that 
the expression for the distinguishing mark $\phi^{sc}_{E}(a,s)$ of 
a Euclidean universe is given by
\begin{equation}  \label{MarkE}
\phi^{sc}_{E}(a,s) =\frac{3}{16\,a^6}\,\,s^2\,(2a-s)^2\,(s+4a)\;,
\end{equation} 
which holds for $s\in (0, 2a]$, and  where the value of the
normalization constant was found to be $\zeta = 9/(16\pi ^2 a^6)$.

Before proceeding to the next class it should be noticed that 
the shape of the signature $\phi^{sc}_{E}(a,s)$ does not 
depend on the value of the radius $a$. Indeed, in terms of a new 
variable $s'= s/a$ the expression~(\ref{MarkE}) can be rewritten 
in the form
\begin{equation}  \label{fdensEa}
\phi^{sc}_{E} (a,s')= \frac{3}{16\,a}\,\,s'^{2}\, 
                                      (2-s')^2 \, (s'+4)\;,
\end{equation} 
which makes clear that for distinct radii $a$ one has dif\/ferent 
constant multiplying factors $3/(16a)$, but without changing the 
functional dependence of $\phi^{sc}_{E}(a,s)$ with $s$.
So, the  shape of the graph of the distinguishing mark function
for Euclidean universes does not depend on the value of the 
radius $a$. 

For hyperbolic universes ($\mathcal{B}_a \subset H^3$) one obtains   
$\mathcal{A}_S(r)=4\pi \sinh^2r$ and $\mathcal{A}_C(a,r,s)=2\pi\sinh s\,
[\,\sinh s-\cosh s \,\coth\, r + \cosh a \,\,\mbox{csch}\,r\,]$.
Again, through the second and third steps of the above-mentioned 
systematic scheme one f\/inds the following expression for the 
distinguishing mark of a simply-connected hyperbolic universe fulf\/illed 
with an uniform distribution of objects:
\begin{equation}  \label{MarkH}
\phi^{sc}_H(a,s)=\frac{8 \sinh^2 s}{(\sinh 2a -2a)^2}\,\,\left[
\,\cosh a \,\,\mbox{sech}(s/2)\,\sinh(a-s/2)\,\,-\,(a-s/2)\,\right]\;, 
\end{equation}  
which holds for $s\in (0,2a]$, and where the value of the 
normalization constant in this case was found to be 
$\zeta = [\,\pi \,(\sinh 2a - 2a)\,]^{-2}\,$. 
 
As for the elliptic universes ($\,\mathcal{B}_a \subset S^3$) 
when the diameter  $2a$ is less than the separation $\pi R$ 
between antipodal points of $S^3$ the above scheme can be 
similarly used. For this case one obtains
$\mathcal{A}_S(r)=4\pi \sin^2 r\,$ and $\,\mathcal{A}_C(a,r,s)= 
2\pi\sin s\,[- \sin s - \cos s \,\cot\,r + \cos a \,\,\csc\,r\,]\,$.
Using these expressions in~(\ref{den}) and following the 
above-outlined general procedure one f\/inds 
\begin{equation}  \label{PartMarkEL}
\phi^{sc}_S(a,s)=\frac{8 \sin^2 s}{(2a-\sin 2a)^2}\,\,\left[\,
(a-s/2)-\cos \,a \,\,\sec (s/2)\,\,\sin(a-s/2) \,\right]\;, 
\end{equation} 
which hold for $2a<\pi$, where we have taken $R=1\,$. The value of 
the normalization constant in this case was found to be 
$\zeta = [\,\pi\,(2a - \sin 2a)\,]^{-2}\,$.

The elliptic universes ($\,\mathcal{B}_a \subset S^3$) for which
$2a>\pi R $ cannot be included in the above general scheme. 
They are trickier to be handled due to the connectivity of the 
spherical space $S^3$ and the additional requirement that $s$ 
must not exceed $\pi R$, which is needed to ensure that one is 
taking the shortest geodesic part between two points of $S^3$. 
For the sake of brevity and completeness we shall present here 
only the f\/inal expression for $\phi^{sc}(a,s)\,$. 
It turns out that the general expression of the distinguishing
mark which holds for all elliptic universes with $a\in(0,\pi]$ and
fulf\/illed with a uniform distribution of objects
is given by
\begin{eqnarray} 
\phi^{sc}_S(a,s)&=& \frac{8\sin^2 s}{(2a-\sin 2a)^2} \,\,\left\{
   2a-\sin 2a-\pi + \Theta(2\pi-2a-s)\,\,\left[\,\sin 2a\,+\,\pi
            \right. \right. \nonumber \\ 
& & \qquad \qquad \qquad \quad
  -\,a-s/2 - \cos a\,\, \sec(s/2)\,\,\sin(a-s/2)\, \left. \right]\, 
                     \left.\right\} \label{MarkEL} \;,
\end{eqnarray}
where $s\in (0,\mbox{min}(2a,\pi)]\,$.  

In what follows we shall present and analyze a few graphs of the 
distinguishing mark for each class or RW simply-connected 
universes.

Figure~2 shows the distinguishing mark $\phi^{sc}_{E}(a,s)$
for a Euclidean universe $\mathcal{B}_a$ with radius $a=0.5\,$.
This marking also gives the probability distribution of the pair 
separation distance $s$ for $s\in (0, 2a]$. 
A close  inspection of this f\/igure reveals that the
most likely separation between two arbitrary pointlike
objects in the Euclidean observed universe $\mathcal{B}_a$ 
is slightly greater than the radius $a$ of the ball. 

Figure~3 shows the distinguishing mark $\phi^{sc}_{H}(a,s)$ 
for three values of the radius $a$. For $a \ll 1$ this function 
behaves approximately as that we have derived for the Euclidean universe 
(f\/igure~2), as one would expect from the beginning. For increasing 
values of the radius $a$ of the observed universe the maximum of the 
mark $\phi^{sc}_{H}(a,s)$ moves towards the greater values of 
$s$. In other words, the most likely value of $s$ increases 
(the maximum of $\phi^{sc}_{H}(a,s)$ takes place later) 
for increasing values of the radius $a\,$.

In f\/igure~4 four graphs of the distinguishing mark 
$\phi^{sc}_{S}(a,s)$ for dif\/ferent values of the 
radius $a$ of the universe are shown.  
For increasing values of $a$ from $0$ to $\pi$ the maximum of
the signature $\phi^{sc}_{S}(a,s)$ moves continuously 
towards the smaller values of $s$. 
Contrarily to the hyperbolic case, here for increasing values of 
the radius $a$ the maximum of $\phi^{sc}_{S}(a,s)$ moves
toward the origin (smaller values of $s\,$). This is also illustrated 
in f\/igure~4, where the most likely value of the separation $s$ 
decreases (the maximum of $\phi^{sc}_{S}(a,s)$ takes place 
earlier) for increasing values of $a\,$.

It also should be mentioned that for an arbitrary radius of curvature 
of the geometry $R$ the expressions for $\,\phi^{sc}_{H}(a,s)\,$ 
and $\,\phi^{sc}_{S}(a,s)\,$ can be obtained by multiplying 
the right-hand side of~(\ref{MarkH}) and~(\ref{MarkEL}) by $1/R$, 
and simultaneously by changing 
$a \rightarrow a/R$ and $s \rightarrow s/R\,$.

In the remainder of this section we will discuss the improvement 
of the method to extract the topological signature 
\begin{equation}  \label{topsig}
\varphi^{mc}(s_i)=(n-1)\,[\,\Phi_{exp}(s_i)
                            -\Phi^{sc}_{exp}(s_i)\,]\;,
\end{equation}
of multiply-connected universes studied in~\cite{GRT99b} (see 
also~\cite{GRT99b}). To this end, the relevant point to be noted 
is that an EPSH $\Phi_{exp}(s_i)$ is essentially a typical PSH from 
which the statistical noise has been withdrawn. Hence we have
\begin{eqnarray}  
\Phi_{exp}(s_i) & = &\Phi(s_i) - \rho^{mc}(s_i) \label{noise1} \;, \\
\Phi^{sc}_{exp}(s_i)&=&\Phi^{sc}(s_i)-\rho^{sc}(s_i) \label{noise2} \; ,
\end{eqnarray}
where $\rho^{mc}(s_i)$ and $\rho^{sc}(s_i)$ represent the statistical 
noises that arise in the corresponding PSH's.
Using now the decompositions~(\ref{noise1}) and~(\ref{noise2}) 
together with~(\ref{topsig}) one readily obtains
\begin{equation}   \label{topsigapx} 
\varphi^{mc}(s_i) = \,(n-1)\,[\,\Phi(s_i) - \Phi^{sc}(s_i) 
                           +\rho^{sc}(s_i)- \rho^{mc}(s_i)\,] \;,
\end{equation}
which clearly gives the topological signature intermixed
with two statistical f\/luctuations.
Now, from equations~(\ref{topsig})~--~(\ref{topsigapx}) it is
clear that one can approach the topological signature of 
multiply-connected universes $\varphi^{mc}(s_i)\,$ by reducing the 
statistical f\/luctuations, i.e. by making $\rho^{mc}(s_i) \to 0$ as 
well as $\rho^{sc}(s_i) \to 0$ through any suitable statistical 
method to lower the noises. 
An improvement of the method devised in~\cite{GRT99b} to 
extract the topological signature $\varphi^{mc}(s_i)$ of multiply%
-connected universes with a uniform distribution of matter 
comes out from the very fact that  having the derived 
expressions~(\ref{MarkE}), (\ref{MarkH}) and~(\ref{MarkEL}) 
one has from the beginning $\rho^{sc}(s_i)=0\,$ for those universes.
Thus, for example, if the MPSH is the technique one uses to reduce 
the statistical f\/luctuations, the topological signature~(\ref{topsig}) 
in these cases reduces to the form  $\varphi^{mc}(s_i) \simeq 
(n-1)\,[\,<\!\Phi(s_i)\!>-\,\,\Phi^{sc}_{exp}(s_i)\,]\,$, 
with the exact expression for $\Phi^{sc}_{exp}(s_i)$ rather than 
the approximate mean $<\!\Phi^{sc}(s_i)\!>\,$. 

%%%
%%%%%%%%
%%%
\section{Concluding remarks}
\label{concl}
\setcounter{equation}{0}
To a certain extent it is well-known that RW geometry~(\ref{RWmetric}) 
does not f\/ix the global shape (topology) of the spacetime, and that 
there is an inf\/inite number of topologically distinct $t=const$ spatial 
sections $M$ for the RW spacetime manifold. Nevertheless, it is often 
(implicitly or explicitly) assumed that the $t=const$ spatial sections 
$M$ of a RW spacetime manifold are one of the following simply-connected 
spaces:  $E^3$ ($k=0$), $S^3$ ($k=1$), or $H^3$ ($k=-1$). 
However, this assumption of simply-connectedness for our three-space has 
not been settled by cosmological observations. As a matter of fact, 
neither the simply nor the multiply-connectedness for the three-space 
where we live has been discarded or conf\/irmed by the available 
astrophysical data. 

The two main approaches to constrain or determine the topology of
the our three-space rely on the existence of multiple (topological)
images of either cosmic objects or spots  of microwave 
background radiation, and thus they aim at non-trivial topology 
of \emph{small} universes --- a possible simply-connectedness 
(trivial topology) of the universe has not been suitably considered 
in these approaches. 

A special method to determine possible non-trivial topologies of RW
universes, and which relies on the existence of multiple images,
was recently discussed in~\cite{GRT99b}. 
There it is suggested that the quantity $\varphi^{mc}(s_i) 
\equiv (n-1)\,[\,\Phi_{exp}(s_i)- \Phi^{sc}_{exp}(s_i)\,]$ is a
suitable measure of the topological signature of the 
multiply-connected RW universes. 
However, $\varphi^{mc}\,(s_i)$ cannot be used as the topological 
signature for simply-connected universes, since it vanishes identically.
This means that if we live in RW simply-connected universe the graphs 
of $\varphi^{mc}(s_i)$ which arise from real (or simulated) catalogs 
will exhibit nothing but statistical noise. 
Thus $\varphi^{mc}(s_i)$ can be used not only to extract the topological 
signature of multiply-connected universes but also to decide between 
multiply or simply-connectedness of the universe, since in this latter 
case it gives rise simply to statistical noise.

One might think at f\/irst sight that  the vanishing of 
$\varphi^{mc}(s_i)\,$ (which means that it gives rise to 
nothing but statistical f\/luctuations) would lead \emph{only} 
to the simply-connectedness without separating  among the three 
possible classes of simply-connected RW universes. In practice, 
though, the vanishing of $\varphi^{mc}(s_i)$ takes place for one 
underlying RW metric used to convert redshift into distance to have 
the pair separations, and thus it also gives the underlying manifold 
of the corresponding simply-connected universe, as there is a clear 
correspondence between geometry and the covering manifold in these cases. 
Nevertheless, since $\varphi^{mc}(s_i)\,$ vanishes identically 
(gives rise to nothing but statistical noise) for all classes of 
RW universes it is not an univocal (or unequivocal) distinguishing 
mark of these universes. Thus we have been led to take the quantity 
$\phi^{sc}(a,s)$ as distinguishing mark of the simply-connected RW 
universes.
Clearly $\phi^{sc}(a,s)\,$: (i) does not vanish; (ii) can be
used to separate the three possible classes of simply-connected
RW universes; and (iii) it is the quantity which really matters (in 
this statistical context) for the simply-connected cases.
Actually $\phi^{sc}(a,s)$ can also be used to distinguish RW 
universes with dif\/ferent radius $a$.
Further, note that since the way one measures the distances varies 
for each constant curvature universe, it was really expected from
the outset that the expression for the distinguishing mark  
$\phi^{sc}(a,s)$ would be distinct for dif\/ferent 
simply-connected universes. 

\begin{sloppypar}
We have also presented the explicit expressions of the distinguishing 
marks $\phi^{sc}_E(a,s)\,$, $\phi^{sc}_H(a,s)\,$ and  
$\phi^{sc}_S(a,s)\,$ for, respectively, Euclidean, hyperbolic 
and elliptic simply-connected RW universes fulf\/illed with an uniform 
distribution of cosmic objects. 
Besides, we have presented and analyzed graphs of this signature 
for the simply-connected RW universes, and discussed the improvement 
that these exact expressions bring to the method to extract the 
topological signature of multiply-connected universes discussed 
in~\cite{GRT99b}.   
\end{sloppypar}

The distinguishing marks for the simply-connected RW universes 
$\phi^{sc}_E(a,s)\,$, $\phi^{sc}_H(a,s)\,$ and $\phi^{sc}_S(a,s)\,$, 
which we have studied in is this work give, in each case, the probability 
distributions of the pair separation $s \in (0,2a]$. If one takes
these probability distributions as \emph{ground} distributions, then the 
topological signature of multiply-connected RW universes $\varphi^{mc}(s_i)$ 
studied in~\cite{GRT99b} can be understood as a measure of the 
deviation between the pair separation probability distribution in the 
multiply-connected cases [given by $\Phi_{exp}(s_i)\,$] and the 
corresponding ground pair separation probability distribution. 
The isometries $g$ of the covering group $\Gamma$ modify the
ground pair separation probability distribution, and the quantity 
$\varphi^{mc}(s_i)$ measures that deviation of topological
origin.

It is worth mentioning that in the cases of multiply-connected universes 
for which the smaller length of the fundamental polyhedron $\mathcal{P}$ 
of $M$ ($\mathcal{P} \subset \widetilde{M}$) is greater than the diameter
$2 R_H$ ($R_H$ is the particle horizon) of the observed universe 
$\mathcal{B}_{R_H} \subset \widetilde{M}$, no multiple images can be 
observed. These multiply-connected universes are therefore indistinguishable 
{}from the simply-connected universes with the same covering space, equal 
radius, and identical distribution of cosmic sources. 
In these multiply-connected cases in which the scale of the 
multiply-connectedness is greater than the radius $R_H$ universe 
$\mathcal{B}_{R_H}$ no sign of the multiply-connectedness will 
arise, and the distinguishing marks we have discussed in this 
work can play a relevant role, when there is a homogeneous 
distribution of matter, of course.

To close this article it is worth mentioning that the ultimate goal
in the statistical approaches to extract the topological signature 
(mark) is the comparison of the graphs (signature or mark) obtained 
either theoretically or from simulated catalogs against similar 
graphs obtained from real catalogs. To do so, one clearly has to 
have either the exact explicit expression or the simulated patterns
of the topological signatures of the possible universes.
The expressions we have found for the distinguishing mark of
simply-connected RW universes with a uniform distribution of matter
can certainly be used in such comparisons. Note, however, that even 
if the universe turns out to be simply-connected one still has to 
face the remaining problem of reducing the noise from just one or 
even a few real catalogs of cosmic sources.

%%%
%%%%%%%%
%%%
\vspace{3mm}
\section*{Acknowledgments}
I thank A.A.F. Teixeira and  G.I. Gomero for stimulating and 
fertile discussions, for the careful reading of the manuscript 
and indication of relevant omissions. I also thank the 
scientif\/ic agency CNPq for financial support.

%%%
%%%%%%%%
%%%
\vspace{3mm}
\section*{Captions for the f\/igures}

\begin{description}
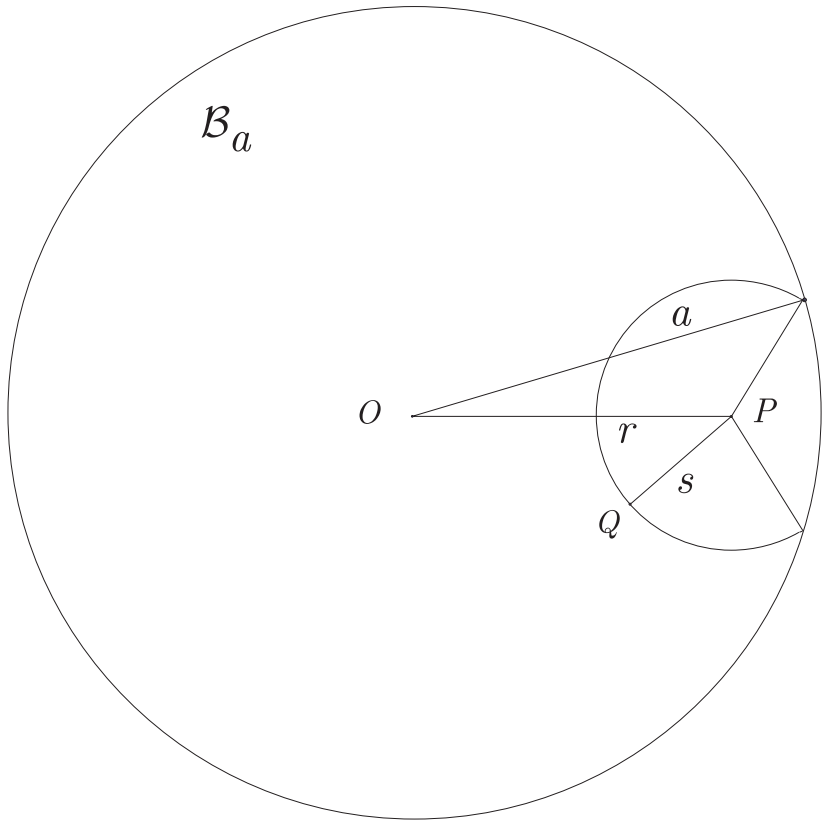
\item[Figure 1.] Two-dimensional schematic f\/igure of observed
universe: spherical ball $\mathcal{B}_a$ of radius $a$, which 
contains the set of the observed images. The circular arc with
center in $P$ represents a spherical calotte (cap)
that changes into a sphere $S^2$ when $r+s \leq a$.

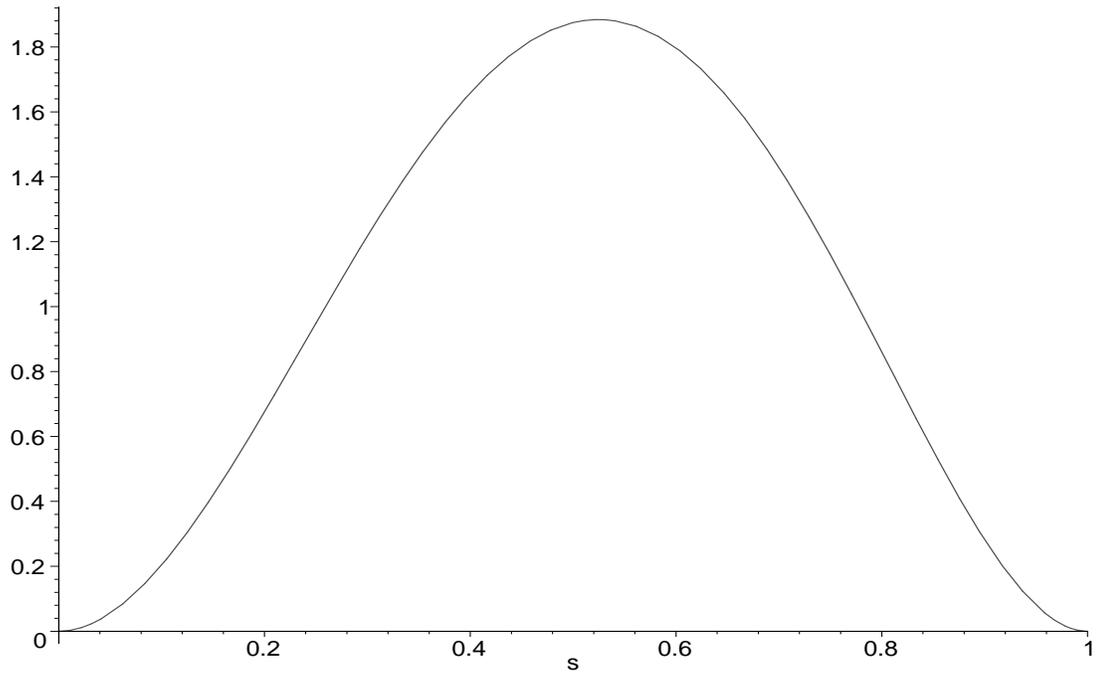
\item[Figure 2.] The distinguishing mark $\phi^{sc}_{E}(a,s)$
for a Euclidean simply-connected universe $\mathcal{B}_a$ for a radius 
$a=0.5\,$. The horizontal axis gives the pair separation $s$ while the
vertical axis gives the normalized number of pairs.
This curve also gives the probability distribution of the 
pair separation distance $s$ in this universe $\mathcal{B}_a$. 
A close inspection reveals that the most likely separation 
between two arbitrary images in a Euclidean universe $\mathcal{B}_a$ 
is slightly greater than the radius $a$ of the universe. 

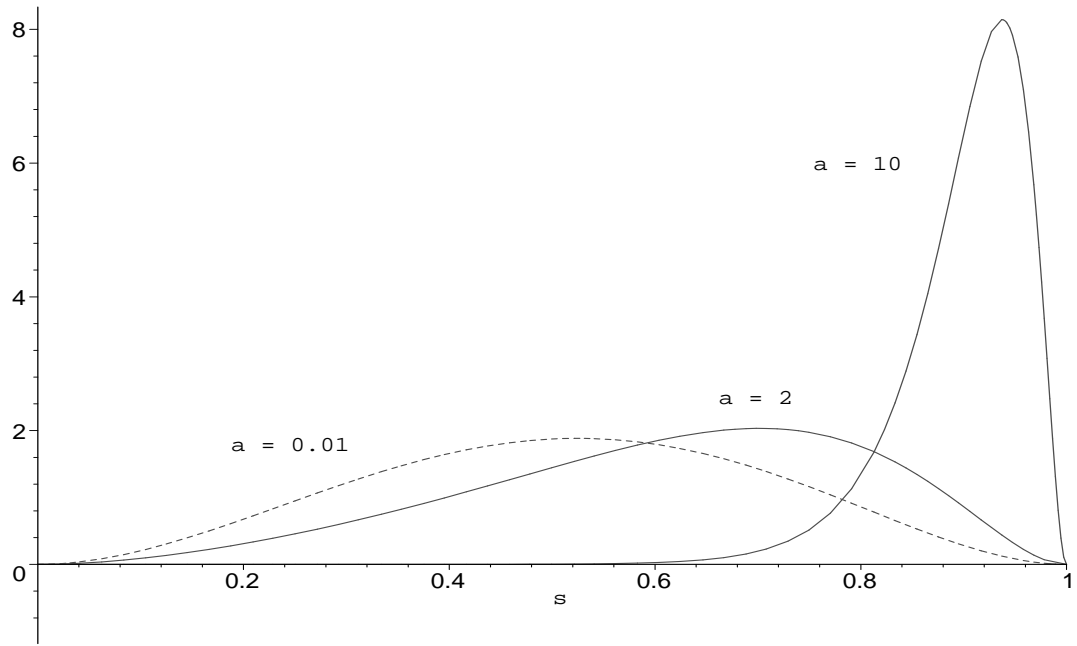
\item[Figure 3.] Graphs of the distinguishing mark
$\phi^{sc}_{H}(a,s)$ of hyperbolic simply-connected universes 
$\mathcal{B}_a$ for three dif\/ferent values of the radius $a$. The 
normalized number of pairs in the vertical axis is given in unit 
of $1/(2a)\,$, while in the horizontal axis the unit of length is 
equal to $2a\,$.
Note that for $a \ll 1$ the signature $\phi^{sc}_H(a,s)$ behaves 
approximately as its Euclidean counterpart $\phi^{sc}_E(a,s)$, 
whose graph is shown in f\/igure~2. For increasing values of the radius 
$a$ of the universe $\mathcal{B}_a$ the maximum of the signature
$\phi^{sc}_{H}(a,s)$ shifts to the right. The most likely 
value of $s$ increases for increasing values of the radius $a\,$, 
and for $a \gg 1$ there is noticeable concentration of large 
values of $s$ near the extreme value $s=2a$. 

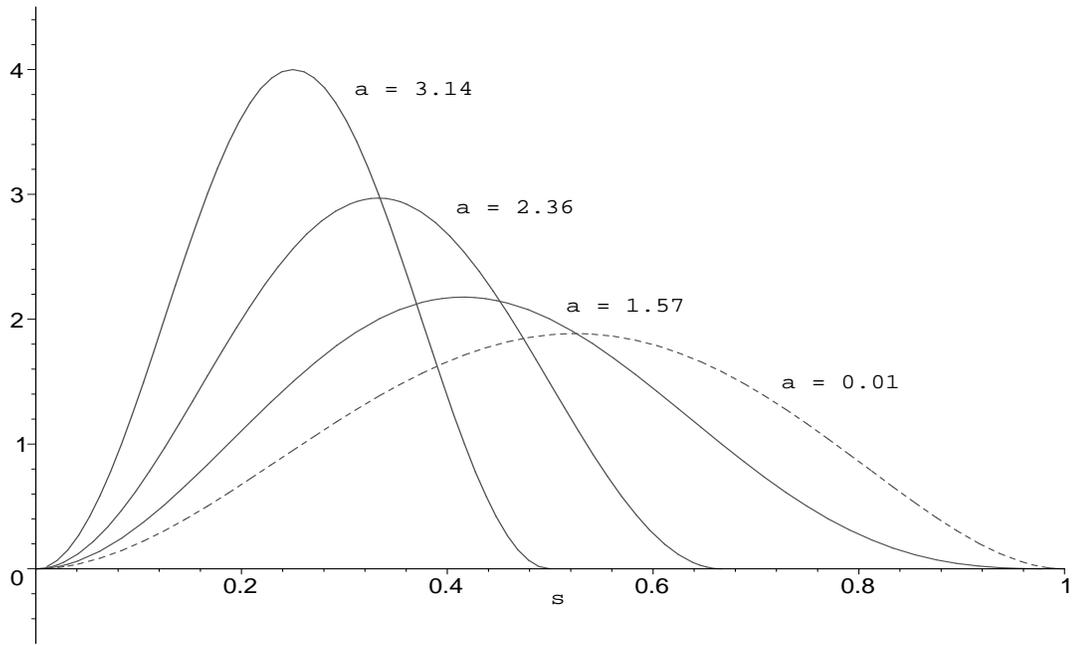
\item[Figure 4.] Graphs of the distinguishing mark
$\phi^{sc}_{S}(a,s)\,$ of elliptic simply-connected universes 
$\mathcal{B}_a$ for four dif\/ferent values of the radius $a$.
The normalized number of pairs in the vertical axis is given in 
unit of $1/(2a)\,$, while in the horizontal axis the unit of length 
is equal to $2a\,$.
For increasing values of $a$ from $0$ to $\pi$ the maximum of the 
signature  $\phi^{sc}_{S}(a,s)$ moves towards the smaller 
values of $s$. This behavior is the opposite of that corresponding
to the hyperbolic case shown in f\/igure~3. The most likely value 
of the separation $s$ decreases for increasing values of $a\,$.

\end{description}

\end{document}